\documentclass[aps,prl,preprint,nopacs,superscriptaddress]{revtex4}

\usepackage{graphicx}
\usepackage{verbatim}
\usepackage{mathrsfs}
\usepackage{color}
\usepackage{amsmath,amsfonts,amssymb}
\usepackage{graphicx}

\def\3{2.8in}    
\def\2{2.5in}
\def\4{3.0in}

\def \beq {\begin{equation}}
\def \eeq {\end{equation}}
\begin{document}

\title{Discovery of Lorentz-violating Weyl fermion semimetal state in LaAlGe materials}
\author{Su-Yang Xu$^*$}\affiliation {Laboratory for Topological Quantum Matter and Spectroscopy (B7), Department of Physics, Princeton University, Princeton, New Jersey 08544, USA}
\author{Nasser Alidoust$^*$}\affiliation {Laboratory for Topological Quantum Matter and Spectroscopy (B7), Department of Physics, Princeton University, Princeton, New Jersey 08544, USA}

\author{Guoqing Chang$^*$}\affiliation{Centre for Advanced 2D Materials and Graphene Research Centre National University of Singapore, 6 Science Drive 2, Singapore 117546}\affiliation{Department of Physics, National University of Singapore, 2 Science Drive 3, Singapore 117542}
\author{Hong Lu$^*$}\affiliation{International Center for Quantum Materials, School of Physics, Peking University, China}
\author{Bahadur Singh\footnote{These authors contributed equally to this work.}}\affiliation{Centre for Advanced 2D Materials and Graphene Research Centre National University of Singapore, 6 Science Drive 2, Singapore 117546}

\author{Ilya Belopolski}\affiliation {Laboratory for Topological Quantum Matter and Spectroscopy (B7), Department of Physics, Princeton University, Princeton, New Jersey 08544, USA}
\author{Daniel S. Sanchez}\affiliation {Laboratory for Topological Quantum Matter and Spectroscopy (B7), Department of Physics, Princeton University, Princeton, New Jersey 08544, USA}

\author{Xiao Zhang}\affiliation{International Center for Quantum Materials, School of Physics, Peking University, China}
\affiliation{Department of Physics, National University of Singapore, 2 Science Drive 3, Singapore 117542}

\author{Guang Bian}\affiliation {Laboratory for Topological Quantum Matter and Spectroscopy (B7), Department of Physics, Princeton University, Princeton, New Jersey 08544, USA}
\author{Hao Zheng}\affiliation {Laboratory for Topological Quantum Matter and Spectroscopy (B7), Department of Physics, Princeton University, Princeton, New Jersey 08544, USA}

\author{Marius-Adrian Husanu}\affiliation {Paul Scherrer Institute, Swiss Light Source, CH-5232 Villigen PSI, Switzerland}
\affiliation{National Institute of Materials Physics, 405A Atomistilor Str. 077125, Magurele, Romania}
\author{Yi Bian}\affiliation{International Center for Quantum Materials, School of Physics, Peking University, China}


\author{Shin-Ming Huang}
\affiliation{Centre for Advanced 2D Materials and Graphene Research Centre National University of Singapore, 6 Science Drive 2, Singapore 117546}
\affiliation{Department of Physics, National University of Singapore, 2 Science Drive 3, Singapore 117542}
\affiliation{Department of Physics, National Sun Yat-Sen University, Kaohsiung 804, Taiwan}
\author{Chuang-Han Hsu}\affiliation{Centre for Advanced 2D Materials and Graphene Research Centre National University of Singapore, 6 Science Drive 2, Singapore 117546}
\affiliation{Department of Physics, National University of Singapore, 2 Science Drive 3, Singapore 117542}
\author{Tay-Rong Chang}
\affiliation{Department of Physics, National Tsing Hua University, Hsinchu 30013, Taiwan}

\author{Horng-Tay Jeng}
\affiliation{Department of Physics, National Tsing Hua University, Hsinchu 30013, Taiwan}
\affiliation{Institute of Physics, Academia Sinica, Taipei 11529, Taiwan}


\author{Arun Bansil}
\affiliation{Department of Physics, Northeastern University, Boston, Massachusetts 02115, USA}
\author{Vladimir N. Strocov}\affiliation {Paul Scherrer Institute, Swiss Light Source, CH-5232 Villigen PSI, Switzerland}

\author{Hsin Lin$^{\dag}$}
\affiliation{Centre for Advanced 2D Materials and Graphene Research Centre National University of Singapore, 6 Science Drive 2, Singapore 117546}
\affiliation{Department of Physics, National University of Singapore, 2 Science Drive 3, Singapore 117542}

\author{Shuang Jia$^{\dag}$}
\affiliation{International Center for Quantum Materials, School of Physics, Peking University, China}\affiliation{Collaborative Innovation Center of Quantum Matter, Beijing,100871, China}

\author{M. Zahid Hasan\footnote{Corresponding authors (emails): nilnish@gmail.com, mzhasan@princeton.edu }}\affiliation {Laboratory for Topological Quantum Matter and Spectroscopy (B7), Department of Physics, Princeton University, Princeton, New Jersey 08544, USA}

\date{\today}

\begin{abstract}

In quantum field theory, Weyl fermions are relativistic particles that travel at the speed of light and strictly obey the celebrated Lorentz symmetry. The low-energy condensed matter analog is the Weyl fermion semimetal state, which are conductors whose electronic excitations mimic the Weyl fermionic equation of motion. Although quasiparticles in condensed matter systems can break Lorentz symmetry, the emergent Weyl fermions in traditional (type-I) Weyl semimetal state, which has now been realized in the TaAs class of materials, still approximately respect Lorentz symmetry. Recently, the so-called type-II Weyl semimetal has been proposed in WTe$_2$ materials (but never observed), where the emergent Weyl quasiparticles break the Lorentz symmetry most dramatically. Despite intense interest and various material suggestions, direct observation and realization for such exotic fermions remain experimentally elusive. In this paper, we present the first experimental discovery and direct observation of the strongly Lorentz-violating Weyl fermions and the type-II Weyl semimetal state in crystalline solid lanthanum aluminum germanide, LaAlGe. Specifically, our systematic bulk-sensitive angle-resolved photoemission spectroscopy (ARPES) data clearly show that the Weyl fermion nodes appear as the touching points between electron and hole pockets, and that the two bands that cross each other forming the Weyl fermion cone have the same sign of velocity along a certain momentum space direction. Our surface-sensitive ARPES data combined with detailed calculations further show evidence for Fermi arc surface states. Our work opens up experimental opportunities for studying novel spectroscopic and transport phenomena due to the type-II Weyl fermions and provides a material platform for testing exotic Lorentz-violating theories beyond the Standard Model in table-top experiments.

[In this paper the term discovery refers to the theoretical and experimental identification of the compounds (LaAlGe) as a class of Lorentz-violating tilted Weyl materials. A detailed theoretical paper describing the topology of the full family of X(Lanthanides)AlGe and their chemical variant materials will follow.]

\vspace{0.6cm}

\end{abstract}
\maketitle

In 1937 physicist Conyers Herring theoretically identified the conditions under which electronic bands in solids have the same energy and momentum by accident in crystals that lack certain symmetries \cite{Herring}. Near these accidental band touching points, the low-energy excitations, or electronic quasiparticles, can be described by an equation that is essentially identical to the 1929 Weyl equation \cite{Weyl}. In recent times, these touching points are referred as Weyl nodes and have been theoretically studied in the context of topological semimetal/metal materials called the Weyl semimetals \cite{Weyl, Grushin, Herring, Volovik2003, Murakami2007, Wan2011, Burkov2011}. The quasiparticles in the vicinity of these nodes are the emergent Weyl fermions. In a Weyl semimetal, the Weyl fermions behave like Berry curvature monopoles or anti-monopoles depending on their respective chirality. The Weyl fermions of opposite chiralities are separated in momentum space and are connected only through the crystal's boundary by an exotic Fermi arc surface state. The presence of parallel electrical and magnetic fields can break the apparent conservation of the chiral charge due to the chiral anomaly, making a Weyl metal, unlike ordinary nonmagnetic metals, more conductive with an increasing magnetic field \cite{nielsen1983adler} and show novel nonlocal transport behavior \cite{Nonlocal}. Because of these novel properties, Weyl semimetals have the potential to be as important as graphene and topological insulators and have attracted worldwide interest.

Lorentz symmetry is the symmetry that requires physical laws to be independent of the orientation or the boost velocity of the laboratory through space \cite{Lorentz_0}. Such a symmetry laid the groundwork of the quantum field theory (QFT) and Einstein's theory of relativity, and has deep connection to the principle of causation and the Charge, Parity, and Time Reversal Symmetry (CPT) symmetry \cite{Lorentz_1, Lorentz_2, Lorentz_3}. The search for Lorentz violation is one of the central themes of modern physics because many of the theories beyond the Standard Model, with the intention of further incorporating gravity, violate Lorentz symmetry. To date, Lorentz symmetry has survived a century of tests and remained to appear exact in high energy physics. For example, the ``real'' Weyl fermion particles travel at the speed of light and therefore strictly obey Lorentz symmetry. Despite its robustness in high-energy particle physics, Lorentz symmetry is not present in low-energy condensed matter physics as slow-moving emergent electronic quasiparticles in crystals generally break Lorentz symmetry. As an example, in 2012, Grushin considered the consequences of a condensed matter realization of the Lorentz violating quantum field theory in the emergent Weyl fermions of a Weyl semimetal \cite{Grushin2}. Despite the fact that Lorentz symmetry is not required in condensed matter systems, the emergent Weyl fermions of a conventional (type-I) Weyl semimetal respect (or at least approximately respect) the Lorentz symmetry. Specifically, the small Lorentz breaking of the type-I Weyl fermions can be adiabatically removed. Therefore, any type-I Weyl fermion is adiabatically and topologically equivalent to the Lorentz invariant Weyl fermion. The type-I Weyl cone arises from the crossing between two bands that have the opposite sign of velocity and the Fermi surface consists of zero-dimensional isolated points if the Fermi level is set at the energy of the Weyl nodes. Type-I Weyl semimetal has been realized in the TaAs class of crystal \cite{Hasan_TaAs, MIT_Weyl, TaAs_Ding}. Fermi arcs and Weyl fermions have been observed by spectroscopic experiments \cite{Hasan_TaAs, MIT_Weyl, TaAs_Ding}, and the negative magnetoresistance due to the chiral anomaly has been reported in the transport behavior of TaAs further confirming its Weyl state \cite{Chiral_anomaly_ChenGF, Chiral_anomaly_Jia}.

Since 2012, several groups theoretically considered the Lorentz violating emergent Weyl fermions in the form of tilted Weyl cones and, as a result, the condensed matter realization of the Lorentz violating quantum field theory in a Weyl semimetal \cite{Grushin, Emil1, Emil2, WT-Weyl, beenakker}. Unlike the type-I case, the type-II Weyl fermions violate Lorentz symmetry so strongly that they cannot be adiabatically connected back \cite{WT-Weyl}. Such a type-II Weyl fermion resulting from extreme tilting of the cone arises from the crossing between two bands that have the same sign of velocity along certain direction, and, in a Fermi surface, the Weyl nodes appear as the touching points between electron and hole pockets \cite{WT-Weyl}. Some version of this extreme case at the vicinity of a type-I and type-II Weyl critical point was briefly discussed in Ref-\cite{Emil1, Emil2} with a commentary in Ref-\cite{beenakker}. This novel Weyl semimetal state has been predicted to show a number of new and novel properties not possible in the regular (type-I) case including an exotic chiral anomaly whose transport response strongly depends on the direction of the electric current, an anti-chiral effect of the chiral Landau level, a novel anomalous Hall effect, and emergent Lorentz invariant properties due to electron-electron interaction \cite{WT-Weyl, Emil1, Emil2, Grushin, QAH, Coulomb}. Despite intense interest and various material suggestions \cite{WT-Weyl, MT-Weyl, MT-Weyl-2, WMTe2, TaS, TIT, WP2}, direct evidence for type-II Weyl semimetal state remain experimentally elusive. In this paper, we present the theoretical identification and the experimental discovery of Lorentz-violating Weyl fermions in the type-II Weyl semimetal state in the crystalline solid LaAlGe.

In this paper, we show the first experimental realization of the type-II Weyl fermions and type-II Weyl semimetal state in the inversion-breaking single crystalline compound lanthanum aluminum germanide, LaAlGe. Using angle-resolved photoemission spectroscopy, we directly observe the type-II Weyl fermions in the bulk band structure of LaAlGe. Remarkably, we find that the type-II Weyl nodes are located essentially at the Fermi level, which strongly suggests that LaAlGe would be an ideal type-II Weyl semimetal for transport experiments. We further observe evidence for the Fermi arc surface states on LaAlGe's (001) surface. All our ARPES data are in excellent agreement with first-principles band structure calculations, providing further support on our conclusion. Our experimental discovery of the type-II Weyl semimetal state in LaAlGe paves the way for experimentally studying novel spectroscopic and transport phenomena due to type-II Weyl fermions and provides a platform for testing exotic Lorentz-violating high energy theories in table-top experiments.

\begin{figure*}[t]
\includegraphics[width=17cm]{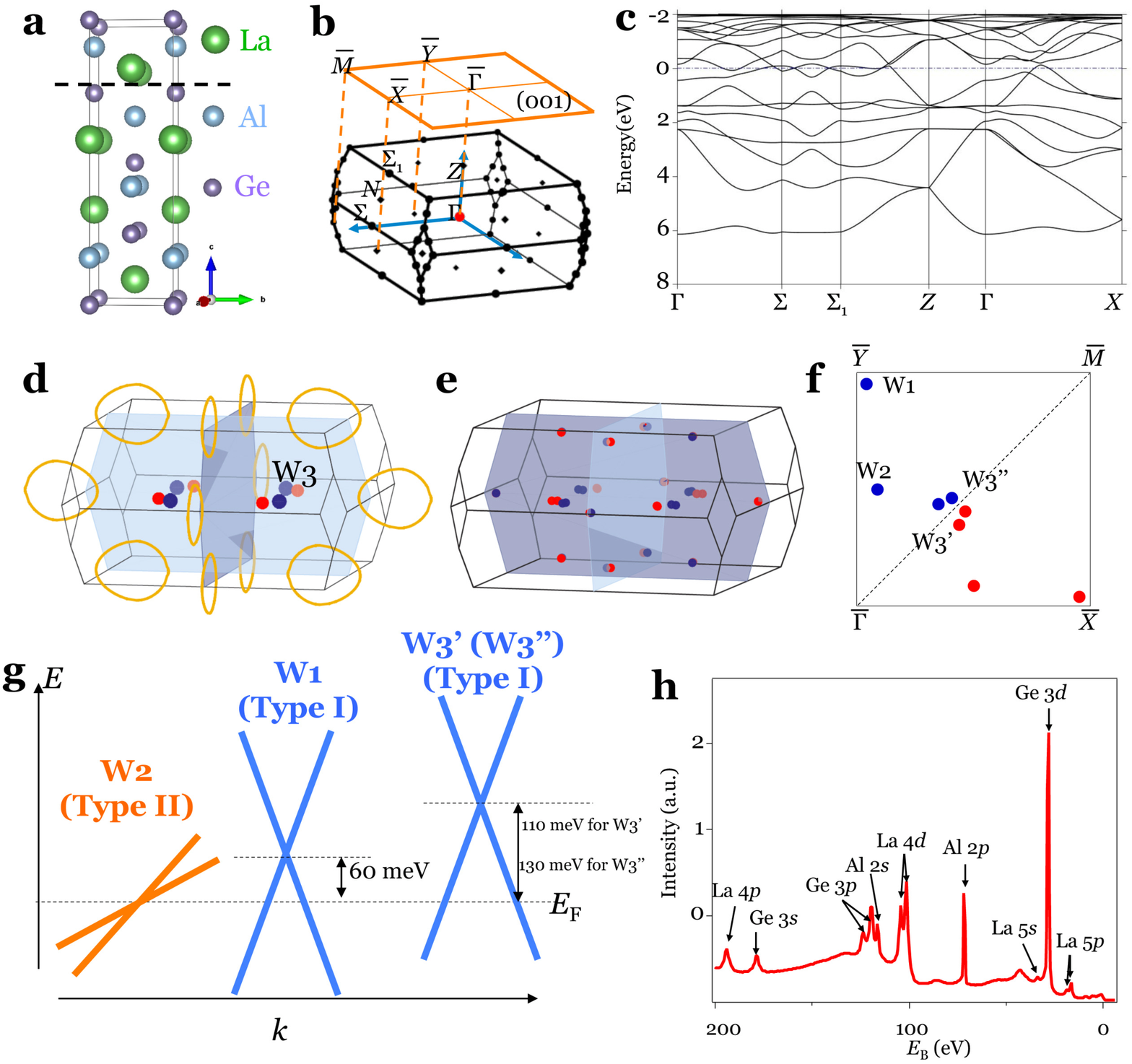}
\caption{
{\bf Topology and Brillouin zone symmetry of LaAlGe.} {\bf a,} Body-centered tetragonal structure of LaAlGe, with space group \textit{I}$4_{1}$\textit{md} (109). The structure consistes of stacks of La, Al, and Ge layers and along the (001) direction each layer consists of only one type of elements, with $a = b \neq c$ {\bf b,} The bulk and (001) surface Brillouin zone of LaAlGe. {\bf c,} First-principles band structure calculations of LaAlGe  along high symmetry directions without spin-orbit coupling (SOC). {\bf d,} Momentum space configuration of the four nodal lines (two on the $k_x = 0$ and two on the $k_y = 0$ mirror planes) denoted by the rings, as well as the 4 spinless pairs of Weyl nodes denoted as W3 on the $k_z = 0$ plane,  in the absence of SOC. Blue and red colors indicate positive and negative chiralities respectively. {\bf e,} Configuration of the 40 Weyl nodes in the Brillouin zone created upon the inclusion of SOC. The nodal lines are gapped out by SOC and 24 Weyl nodes emerge in the vicinity of the nodal lines.}
\label{Lattice}
\end{figure*}

LaAlGe crystallizes in a body-centered tetragonal Bravais lattice, space group $I4_1md$ (109), point group $C_{4v}$. The lattice constants according to previous diffraction experiments are $a=b=4.336$ $\textrm{\AA}$ and $c= 14.828$ $\textrm{\AA}$ \cite{LAG_Structure}. The basis consists of two La atoms, two Al and two Ge atoms, as shown in Fig.~\ref{Lattice}\textbf{a}. In this crystal structure, along the (001) direction, each atomic layer consists of only one type of elements and is shifted relative to the layer below by half a lattice constant in either the $x$ or $y$ direction. This shift gives the lattice a screw pattern along the $z$ direction, which leads to a non-symmorphic $C_4$ rotation symmetry that includes a translation along the $z$-direction by $c/2$. Crucially, LaAlGe lacks space inversion symmetry. The breaking of space inversion symmetry is an important condition for the realization of a Weyl semimetal because in the absence of inversion symmetry all bands are generically singly-degenerate. Fig.~\ref{Lattice}\textbf{c} shows the calculated bulk band structure along high symmetry lines without spin-orbit coupling. It can be seen that the conduction and valence bands cross each other along the $\Gamma-{\Sigma}-{\Sigma_1}$ path, demonstrating a semimetal groundstate. The momentum space configurations of the band crossings are shown in Figs.~\ref{Lattice}\textbf{d,e}. In the absence of spin-orbit coupling (Fig.~\ref{Lattice}\textbf{d}), the crossing between conduction and valence bands yields four closed loops, nodal lines, on the $k_x=0$ and $k_y=0$ mirror planes and also 4 pairs of (spinless) Weyl nodes on the $k_z=0$ plane, which are denoted as W3 in Fig.~\ref{Lattice}\textbf{d}. Upon the inclusion of the spin-orbit coupling (Fig.~\ref{Lattice}\textbf{e}), the nodal lines are gapped out and 24 Weyl nodes emerge in the vicinity, which is similar to the case of TaAs \cite{Hasan_TaAs}. We refer to the 8 Weyl nodes located on the $k_z=0$ plane as W1 and the remaining 16 Weyl nodes away from this plane as W2. Moreover, when spin-orbit coupling is taken into account each W3 (spinless) Weyl node splits into two (spinful) Weyl nodes of the same chirality, which we call W3' and W3''. Hence, in total there are 40 Weyl nodes in a Brillouin zone (BZ) as shown in Fig.~\ref{Lattice}\textbf{e}. The projection of the Weyl nodes on the (001) surface BZ is shown in Fig.~\ref{Lattice}\textbf{f}. All W1 Weyl nodes project as single Weyl nodes in the close vicinity of the surface BZ edge $\bar{X}$ and $\bar{Y}$ points. Two W2 Weyl nodes of the same chiral charge project onto the same point near the midpoint of the $\bar{\Gamma}-\bar{X}$ and $\bar{\Gamma}-\bar{Y}$ lines. The W3' and W3'' are projected near the midpoint of the $\bar{\Gamma}-\bar{M}$ (diagonal) lines. As shown by the schematic in Fig.~\ref{Lattice}\textbf{g}, which is drawn according to our calculation results, W2 Weyl fermions are of type-II whereas W1 and W3 Weyl fermions are of type-I. Very importantly, all W1, W3', and W3'' nodes are far away from the Fermi level in energy. Specifically, they are $\sim60$ meV or $\sim120$ meV above the Fermi level (Fig.~\ref{Lattice}\textbf{g}). As a result, the Fermi surface pockets that arise from nearby pairs of W1 (or W3', W3'') Weyl cones of opposite chirality will merge into each other. In other words, at the Fermi level the Fermi surface pockets from W1 (or W3', W3'') Weyl cones will not carry any distinct chiral charge and become irrelevant to the low-energy physics near the Fermi level. By contrast, W2 nodes are located almost exactly at the Fermi level. Therefore, at the native chemical potential, LaAlGe is purely a type-II Weyl semimetal. Fig.~\ref{Lattice}\textbf{h} shows the angle-integrated photoemission spectrum of LaAlGe over a wide range of binding energy. We identify the La $4p, 4d, 5s$, Al $2s, 2p$ and Ge $3s, 3p, 3d$ core levels, which confirms the correct chemical composition of our samples.

Utilizing the combination of the soft X-ray (SX) angle-resolved photoemission spectroscopy (ARPES) and vacuum ultraviolet (low-energy) ARPES, we systematically and differentially study the electronic structure of the bulk and the surface of LaAlGe. Our SX-ARPES measurements, which are mainly bulk sensitive and deliver high definition of electron momentum in three-dimensional k-space \cite{vladimir} , can uniquely reveal the three-dimensional linearly dispersive bulk Weyl cones and Weyl node. The bulk-sensitive ARPES experiment has been performed at the SX-ARPES endstation of the ADRESS beamline at the Swiss Light Source. On the other hand, our ultraviolet (low-energy) ARPES measurements, which are highly surface sensitive, can detect the Fermi arc surface states. This systematic methodology has been proven crucial in the experimental identification of the type-I Weyl semimetal state in TaAs \cite{Hasan_TaAs}. Here, for the first time, we apply this methodology to a type-II Weyl semimetal candidate in an attempt to discover the first real example of type-II Weyl fermions in LaAlGe.

We focus on our measurements of the bulk Weyl cones using SX-ARPES since the key distinction between type-I and type-II lies in the Weyl fermions in the bulk rather than the Fermi arcs on the surface. Specifically, the experimental demonstration of the type-II Weyl fermion consists of two critical pieces of evidence: (1) In a constant energy contour, a type-II Weyl fermion node reveals itself as the touching point between electron and hole pockets. (2) In an energy-momentum ($E-k$) dispersion cut, a type-II Weyl fermion manifests as the crossing between two bands with the same sign of velocity.

Fig.~\ref{Bulk}\textbf{a} shows the measured $k_z-k_x$ Fermi surface contour of the bulk bands on the $k_y=0$ plane. We observe large identical contours enclosing the $\Sigma$ and $\Sigma_1$ points. These large contours are ``remnant'' of the nodal-lines, which are gapped by spin-orbit coupling. In addition, we observe two small contours on the left- and right-hand sides of the big contour. The measured $k_z-k_x$  Fermi surface is in agreement with the first-principle calculated results. Most importantly, the measurements shown in Fig.~\ref{Bulk}\textbf{a} clearly prove that we are measuring the bulk band structure by SX-ARPES because surface states should have no $k_z$ dispersion. After confirming this, we choose an incident photon energy (i.e. a $k_z$ value) that corresponds to the W2 Weyl nodes. Fig.~\ref{Bulk}\textbf{c} shows the ARPES measured $k_x-k_y$ Fermi surface at the $k_z$ value that corresponds to the W2 Weyl nodes over a wide $k$-range. It can be seen that the Fermi surface consists of small pockets along the $k_x$ and $k_y$ axes. Fig.~\ref{Bulk}\textbf{d} shows a high-resolution zoomed-in Fermi surface map of the region highlighted by the green box in Fig.~\ref{Bulk}\textbf{c}. From Fig.~\ref{Bulk}\textbf{d}, we clearly identify two Fermi pockets, which touch each other at two points located on the opposite sides of the $k_y$ axis. Fig.~\ref{Bulk}\textbf{f} shows the constant energy contour at $E_{\textrm{B}}=25$ meV of the same $k$-region. We see that the upper pocket expands whereas the lower pocket shrinks upon changing the binding energy from $E_{\textrm{B}}=0$ to $E_{\textrm{B}}=25$ meV. This demonstrates that the upper pocket is hole-like whereas the lower one is electron-like. Therefore, the two touching points observed in Fig.~\ref{Bulk}\textbf{d} are touching points between electron and hole pockets. These systematic measurements of the constant energy contours in Figs.~\ref{Bulk}\textbf{d, f}, which are in good agreement with the corresponding first-principles calculations (Fig.~\ref{Bulk}\textbf{e, g}), provide the first critical piece of evidence.

We also show $k_x-k_y$ Fermi surface data at a different photon energy, which corresponds to the $k_z$ value of the W1, W3' and W3'' Weyl nodes ($k_z$=0). Figs.~\ref{Bulk}\textbf{h, i} respectively show the calculated and measured Fermi surface at $k_z=0$. We observe three types of pockets, a small pocket near each BZ boundary, a trapezoid-shaped pocket along each diagonal ($45^{\circ}$) direction, and a circular pocket in-between two adjacent trapezoid pockets. The data (Fig.~\ref{Bulk}\textbf{i}) and calculation (Fig.~\ref{Bulk}\textbf{h}) are in good agreement. The small pocket near the boundary arises from a pair of nearby W1 Weyl fermions, and the trapezoid-shaped pocket arises from pairs of W3' and W3'' Weyl fermions. Since the energy of the W1, W3' and W3'' Weyl nodes are far above the Fermi level, the constant energy contour at the Fermi level already merges into a single pocket and hence does not carry net chiral charge. This data shows that the W1 and W3 are indeed far away from Fermi level in energy and hence they are irrelevant to the low energy physics of LaAlGe.

We now study the energy-momentum dispersion away from the W2 Weyl node along different momentum space cut-directions. In Fig.~\ref{Bulk_disp}\textbf{b} we present the measured $E-k_y$ dispersion map along the Cut $y$ direction indicated in Fig.~\ref{Bulk_disp}\textbf{a}. Here, we clearly resolve two linearly-dispersing hole-like and electron-like bands touching each other upon crossing the Fermi level, in excellent agreement with our first-principles calculations shown in Fig.~\ref{Bulk_disp}\textbf{c}. Cutting along Cut $x$ denoted in Fig.~\ref{Bulk_disp}\textbf{a}, we observe a pair of these band touchings occurring on the opposite sides of the $k_x = 0$ line, again consistent with our calculations. Furthermore, the linear dispersion along the out-of-plane direction for the W2 Weyl nodes is shown by our data in Fig.~\ref{Bulk_disp}\textbf{f, g}. This clearly demonstrates the observation of Weyl cones in LaAlGe. However, the Weyl cones observed here are in clear contrast to those observed in the originally discovered Weyl semimetal TaAs and its related materials. To elaborate on this, we compare these two types of Weyl fermions in Fig.~\ref{Bulk_disp}\textbf{h, i}. In the TaAs Weyl semimetal and other related compounds in that class of materials (such as NbAs, TaP, and NbP), the bulk conduction band and valence band touch each other in a manner that the former is located entirely above a constant energy plane cutting across the Weyl node parallel to the $k_x - k_y$ plane, and the latter entirely below this plane, as shown in Fig.~\ref{Bulk_disp}\textbf{i}. However, in LaAlGe such plane intersects the bulk conduction and valence bands, resulting to the observation that hole-like and electron-like bands cross each other with the same sign of Fermi velocity to form the type-II Weyl node. This signature differentiates the novel type-II Weyl fermions discovered here in LaAlGe from the type-I Weyl fermions observed previously in materials such as TaAs. Figs.~\ref{Bulk_disp}\textbf{j, k} elucidate this difference by comparing the $E - k_y$ dispersion maps of these two compounds. Crucially, the cut along $k_y$ direction in Figs.~\ref{Bulk_disp}\textbf{b,j} provide the second critical piece of evidence that in an energy-momentum ($E-k$) dispersion cut, a type-II Weyl fermion manifests as the crossing between two bands with the same sign of velocity. Therefore, these systematic data in Figs. 2 and 3 taken collectively clearly prove the type-II Weyl fermions and the type-II Weyl semimetal state in LaAlGe.

After demonstrating the type-II Weyl fermions in the bulk, we study the surface electronic structure by low-energy ARPES. Fig.~\ref{SS}\textbf{b} shows measured Fermi surface and constant energy contour data at different binding energies of the (001) surface. We identify the following features in the Fermi surface: (1) We observe a big contour centered at the $\bar{\Gamma}$ point; (2) We observe a tadpole-shaped feature along each $\bar{\Gamma}-\bar{X}(\bar{Y})$ line. The head of the tadpole along the $\bar{\Gamma}-\bar{X}$ direction is truncated by the big circle at the center; (3) We observe two small circular contours in the vicinity of each $\bar{X}$ point; (4) We observe an extended butterfly-shaped contour centered at the $\bar{Y}$ point. By comparing our ARPES data with the calculated surface Fermi surface of different atomic terminations, we are able to determine that the experimental surface termination is the Ge termination between the Ge/La interface (as shown by the dotted line in Fig.~\ref{Lattice}\textbf{a}). On this termination, we find a good agreement between the ARPES measured (Fig.~\ref{SS}\textbf{b}) and the calculated Fermi surfaces (Fig.~\ref{SS}\textbf{a}). Specifically, all the features found in the ARPES data are also seen in the calculation. Before going into the further details, we notice that the Fermi surface strongly breaks $C_4$ symmetry. In the LaAlGe crystal structure, the $C_{4z}$ rotational symmetry is implemented as a screw axis that sends the crystal back into itself after a $C_4$ rotation and a translation by $c/2$ along the rotation axis. Therefore, while the bulk is $C_4$ symmetric (which we indeed found in the data in Figs. 2 and 3), the (001) surface is expected to break the $C_4$ symmetry. As a result, the observed $C_4$ breaking data in Fig.~\ref{SS}\textbf{b} serves as a compelling evidence that we are indeed measuring the surface band structure by low-energy ARPES.

We now focus more closely on selective $k$-regions. In Fig.~\ref{SS}\textbf{d}, we show the high-resolution zoomed-in map near the head of the tadpole feature along the $\bar{\Gamma}-\bar{X}$ direction. This region roughly corresponds to the location at which the W2 Weyl nodes are projected. In Fig.~\ref{SS}\textbf{e}, we superimpose the low-energy surface state Fermi surface (Fig.~\ref{SS}\textbf{d}) and the high energy bulk Fermi surface near W2 (Fig.~\ref{Bulk}\textbf{d}) to scale. We find that the Weyl nodes are not projected onto the tadpole feature we observed. Rather, they fall inside the head of the tadpole. In Fig.~\ref{SS}\textbf{c}, we show the zoomed-in calculation over the same region, where we found a similar scenario. Consequently, the tadpole features are not connected to the Weyl nodes and are hence not Fermi arcs. In the calculated Fermi surface (Fig.~\ref{SS}\textbf{c}), the Fermi arc-like surface states are the very faint features inside the head of the tadpole as they are seen to connect onto the Weyl nodes. We note that the surface calculation considers spectral weight from the top unit cell ($14.828$ $\textrm{\AA}$) of the surface, yet the Fermi arc-like surface states are very weak. Thus these states are mostly localized in the second unit cell or even deeper (see SI). On the other hand, the penetration depth of low energy ($\sim50$ eV) photons is only $\sim5$ $\textrm{\AA}$, which is even smaller than the thickness of the top unit cell. This explains why these faint features are absent in our ARPES data. We elaborate why we call these weak features as Fermi arc-\textit{like} surface states. As we systematically explained in our previous works \cite{TaP_Hasan, NbP_ARPES, Xu science 2014}, in order to show the existence of Fermi arcs in a topological sense, one needs to count the net number of chiral edge state connecting the bulk band gap along a $k$-space loop that encloses a projected Weyl node. However, in the case of LaAlGe, at the (001) surface, the projections of the W2 Weyl nodes coincide with another trivial bulk band pocket at $k_z=0$ (Fig.~\ref{SS}\textbf{f}). Therefore, the projected band structure along the cut that connects the nearby W2 Weyl nodes (noted by the white dotted line in Fig.~\ref{SS}\textbf{c}) is gapless (the left panel of Fig.~\ref{SS}\textbf{f}). This means that any $k$-space loop that encloses a single W2 Weyl node (such as the green loop in Fig.~\ref{SS}\textbf{c}) will not have a fully gapped bulk band structure, and thus, strictly speaking, identifying Fermi arcs associated with W2 (even based on the theoretical calculated band structure) is not possible in a topological sense. Hence, we call these weak features that connect the Weyl nodes as Fermi arc-like surface states. Finally, we study the butterfly-shaped surface states around the $\bar{Y}$ point in Fig.~\ref{SS}\textbf{g}. We see that these butterfly-shaped surface states embrace the projected W3' and W3'' Weyl nodes. In Fig.~\ref{SS}\textbf{h}, we show the calculated constant energy contour at the energy of the W3'' Weyl nodes. We find that, at the energy of the W3'' Weyl nodes ($E_{\textrm{B}}=130$ meV), the butterfly-shaped surface states become the Fermi arcs that directly are terminated onto the W3'' Weyl nodes. A schematic illustration of the Fermi arc connectivity for the W3'' Weyl nodes is shown in Fig.~\ref{SS}\textbf{i}. At the Fermi level, bulk bands from nearby W3' and W3'' already merge as a single pocket. Therefore, the observed butterfly-shaped surface states can be viewed as a Fermi arc-derived surface states associated with W3' and W3'' Weyl nodes. Although the surface states of LaAlGe are quite complicated and a direct demonstration of Fermi arcs based on ARPES data alone is challenging, our systematic studies provide evidence for Fermi arcs based on the good agreement between our data and calculation.

We discuss a few important aspects in connection to our data and experimental observation of the type-II Weyl semimetal state in LaAlGe. First, prior to our discovery, there have been a number of materials theoretically predicted as candidates of type-II Weyl semimetals, including WTe$_2$ \cite{WT-Weyl}, W$_{1-x}$Mo$_x$Te$_2$ \cite{MT-Weyl, MT-Weyl-2, WMTe2}, Ta$_3$S$_2$ \cite{TaS}, TaIrTe$_4$ \cite{TIT} and WP$_2$ \cite{WP2}. Except for the W$_{1-x}$Mo$_x$Te$_2$, all other candidates remain completely experimentally untested. As for the the W$_{1-x}$Mo$_x$Te$_2$, although there have been a number of ARPES experiments \cite{WT-ARPES-1, WT-ARPES-2, WT-ARPES-3, WT-ARPES-4, WT-ARPES-5}, direct and clear evidence for the type-II Weyl fermions remain lacking. This is because of two important factors. (1) The separation of Weyl nodes in pure WTe$_2$ ($x=0$) is extremely small beyond experimental resolution. (2) The predicted Weyl nodes in W$_{1-x}$Mo$_x$Te$_2$ are far above the Fermi level in energy ($\sim50$ meV above $E_{\textrm{F}}$), which cannot be accessed by normal photoemission as ARPES only probes the band structure below $E_{\textrm{F}}$. One recent study has utilized time-resolved (pump-probe) ARPES to access the unoccupied states of W$_{1-x}$Mo$_x$Te$_2$ \cite{WT-ARPES-4}. However, the incident photon energy for time-resolved ARPES is 6 eV (low energy). As we have established in here and also in the TaAs class of materials \cite{Hasan_TaAs, TaP_Hasan}, low energy ARPES is mostly sensitive to the surface state not the bulk bands. Hence, the time-resolved ARPES can, at most, provide evidence for the Fermi arcs in W$_{1-x}$Mo$_x$Te$_2$ but not the Weyl nodes. However, the key distinction between type-I and type-II is in the bulk Weyl fermion not the Fermi arcs. Therefore, to date, direct observation of the type-II Weyl state in W$_{1-x}$Mo$_x$Te$_2$ remain unestablished. Furthermore, we note that because the predicted Weyl nodes in W$_{1-x}$Mo$_x$Te$_2$ are far above $E_{\textrm{F}}$ in energy, these Weyl nodes are therefore irrelevant to low energy physics and phenomena, for example, in transport experiments. By contrast, our systematic ARPES data here provide the first clear and direct observation of the type-II Weyl fermions in LaAlGe in real materials. Equally importantly, in LaAlGe, only the type-II Weyl fermions (W2s) are relevant to low-energy physics. In order words, our data show that LaAlGe is a pure type-II Weyl semimetal in transport experiments. We elaborate on why the other Weyl nodes (W1, W3' and W3'') do not matter in low-energy physics. For example, if we set the energy at W3'' Weyl node ($130$ meV above $E_{\textrm{F}}$), then each W3'' Weyl fermion cone will contribute a point-like constant energy contour as they are type-I. These points carry distinct chiral charges. As we vary the energy away from the W3'' Weyl node, then the constant energy contour from each W3'' Weyl node will become a sphere. At energies close to the  W3'' Weyl node, the spheres are still separated in $k$ space and hence still carry distinct chiral charge. However, at energies far away from the W3'' Weyl node, such as the Fermi level, then the constant energy contours from nearby W3'' Weyl nodes will merge and form a single pocket. In that case, the merged Fermi pocket would not carry distinct chiral charge and hence will not contribute to any Weyl physics (such as the chiral anomaly) in transport. Finally, we suggest that by chemically doping the LaAlGe, one could in-principle engineer the energy position of the chemical potential. For example, both La$_{1-x}$Ce$_x$AlGe and LaAl$_{1-x}$Ge$_{1+x}$ may electron dope the system. By electron doping the LaAlGe system, we may be able to tune the chemical potential in a controlled manner to be at the different Weyl nodes, which can realize a novel transition from type-II Weyl semimetal to a type-I Weyl semimetal in a single system.

In summary, using a combination of bulk-sensitive angle-resolved photoemission spectroscopy at soft-X-ray excitation energies with surface-sensitive photoemission spectroscopy at low energies, we have demonstrated the first experimental realization of the type-II Weyl fermions and type-II Weyl semimetal state. Previous theoretical predictions (WTe$_2$, see \cite{WT-Weyl}, for example) were not ideal for experiments (similar to many predictions of type-I Weyl materials over many years) therefore we first had to theoretically predict and identify a new compound (LaAlGe class of materials reported here) that works for ARPES experiments to discover the type-II Weyl fermions in experiments which is similar to the case for our discovery of TaAs class materials (type-I) previously (many theoretical predictions of type-I Weyl existed for many years but those were not experimentally feasible). Here, in the inversion-breaking lanthanum aluminum germanide (LaAlGe) we have directly observed this novel type of Weyl fermions by mapping the bulk band topology of this compound. Our experimental ARPES results are in excellent agreement with our theoretical first-principles calculations and topological band theory, providing further support in establishing this compound as the first experimentally discovered type-II Weyl semimetal. Notably, the type-II Weyl nodes in this compound are intrinsically located almost precisely at the Fermi level. This has strong implications for the advantages of LaAlGe type-II Weyl semimetal for transport experiments. Moreover, we have observed evidence for the Fermi arc surface states on the (001) surface of this compound. Our experimental discovery of the type-II Weyl semimetal state in LaAlGe paves the way for studying and examining the analogs of strongly Lorentz-violating phenomena in theories of high-energy physics in readily available and feasible experiments. Moreover, our discovery provides an ideal platform for experimentally studying novel spectroscopic, optical, and transport phenomena that may emerge from the type-II Weyl fermions in real materials.

\addtocounter{figure}{-1}
\begin{figure*}[t!]
\caption{In addition, each spinless W3 Weyl node split into two spinful Weyl nodes of the same chirality, which we denote as W3' and W3''. Therefore, there are in total 40 Weyl nodes. For the 24 Weyl nodes that emerge from the gapping of the nodal line, we call the 8 Weyl nodes that are near the boundaries of $k_z=0$ plane as W1 and the other 16 that are away from the $k_z=0$ plane as W2. {\bf f,} Projection of the Weyl nodes on the (001) surface Brillouin zone in one quadrant of the Brillouin zone. {\bf g,} Schematics comparing the three types of Weyl nodes appearing upon the inclusion of SOC. The W2 nodes are type-II Weyl nodes, whereas the W1, W3', and W3'' nodes are type-I. W2 Weyl nodes are located almost exactly at the Fermi level, while W2. W3', and W3'' Weyl nodes are about 60 meV, 110, and 130 meV above the Fermi level, respectively. {\bf h,} Core-level measurement of the studied samples, which clearly shows the expected La, Al, and Ge peaks.}
\end{figure*}

\begin{figure*}[t]
\includegraphics[width=17cm]{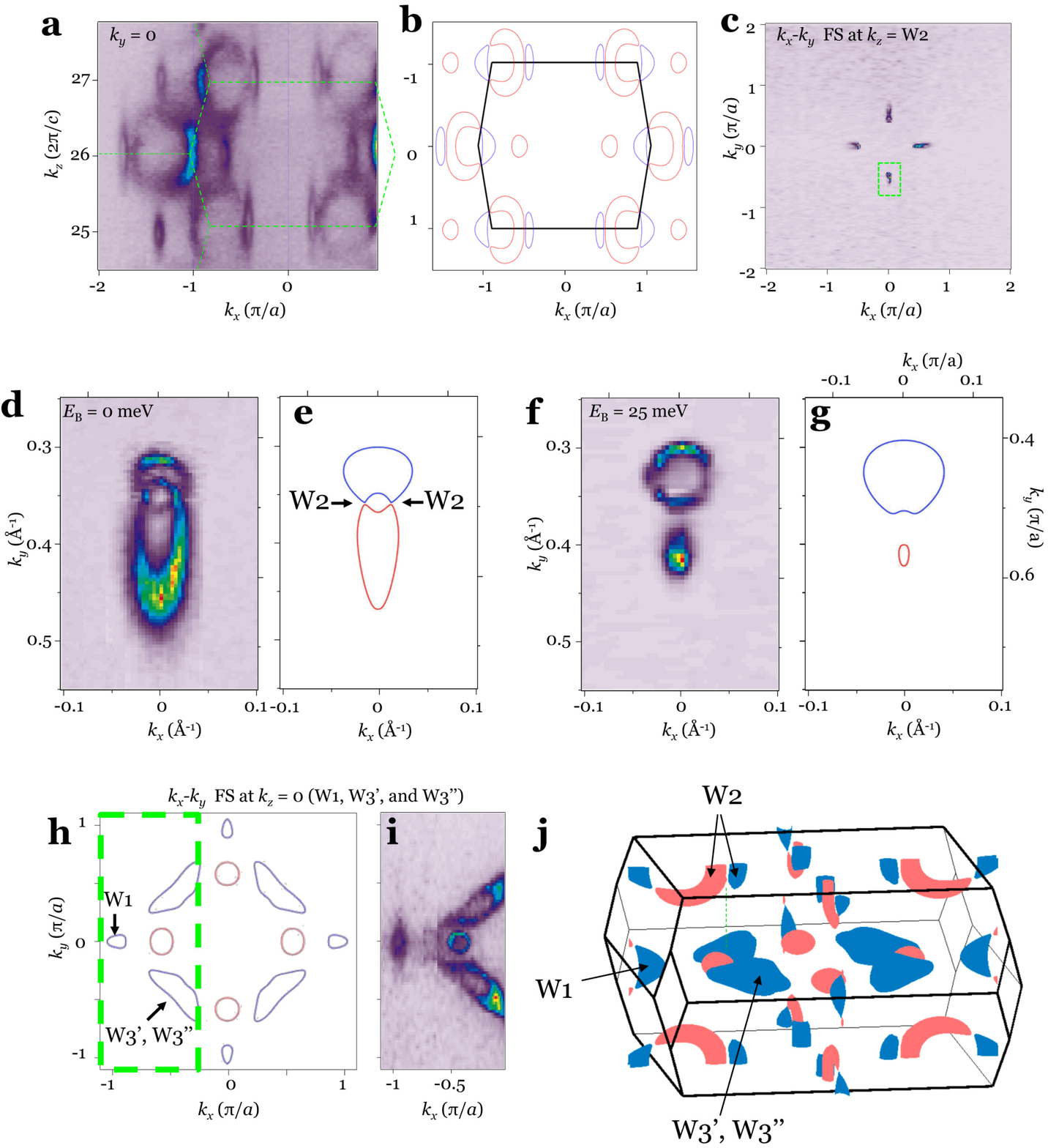}
\caption{
{\bf Observation of bulk Weyl nodes in LaAlGe.} {\bf a,} High-energy soft X-ray angle-resolved photoemission spectroscopy (SX-ARPES) measurement and {\bf b,} first-principles calculation of $k_z-k_x$ Fermi surface maps at $k_y = 0$, respectively. The dashed green hexagon in {\bf a} and the black hexagon in {\bf b} correspond to the first Brillouin zone. The photon energies corresponding to the $k_z$ values in {\bf a} are between 320 eV to 600 eV. {\bf c,} SX-ARPES-measured $k_x-k_y$ Fermi surface map at  $k_z$ corresponding to the W2 Weyl nodes.}
\label{Bulk}
\end{figure*}
\addtocounter{figure}{-1}
\begin{figure*}[t!]
\caption{{\bf d,} Zoomed in version of the measured and {\bf e,} calculated  $k_x-k_y$ Fermi surface maps in the region marked by the green rectangle in {\bf c}. In the first-principles calculations, blue lines correspond to hole-like pockets, whereas red lines indicate electron-like pockets. {\bf f, g,} Same as {\bf d} and {\bf e} for a constant energy contour at 25 meV below the Fermi level. {\bf h,} First-principles calculated $k_x-k_y$ Fermi surface map at $k_z = 0$ (location of W1, W3', and W3'' Weyl nodes), which shows the shapes of the pockets resulting from W1, W3', and W3''. The photon energy used in the measurements presented in {\bf c, d, f} is 542 eV. {\bf i,} SX-ARPES-measured $k_x-k_y$ Fermi surface map at $k_z = 0$ in the region of the Brillouin zone denoted by the green dashed rectangle in {\bf h.} This experimentally observed Fermi surface map is in good qualitative agreement with the calculated results presented in {\bf h.} The photon energy used here is 478 eV. {\bf j,} Configuration of the pockets created by the type-II W2 Weyl cones and the type-I W1, W3', and W3'' Weyl cones in the first Brillouin zone. Pink pockets correspond to electron-like, whereas blue pockets correspond to hole-like pockets.}
\end{figure*}

\begin{figure*}[t]
\includegraphics[width=17cm]{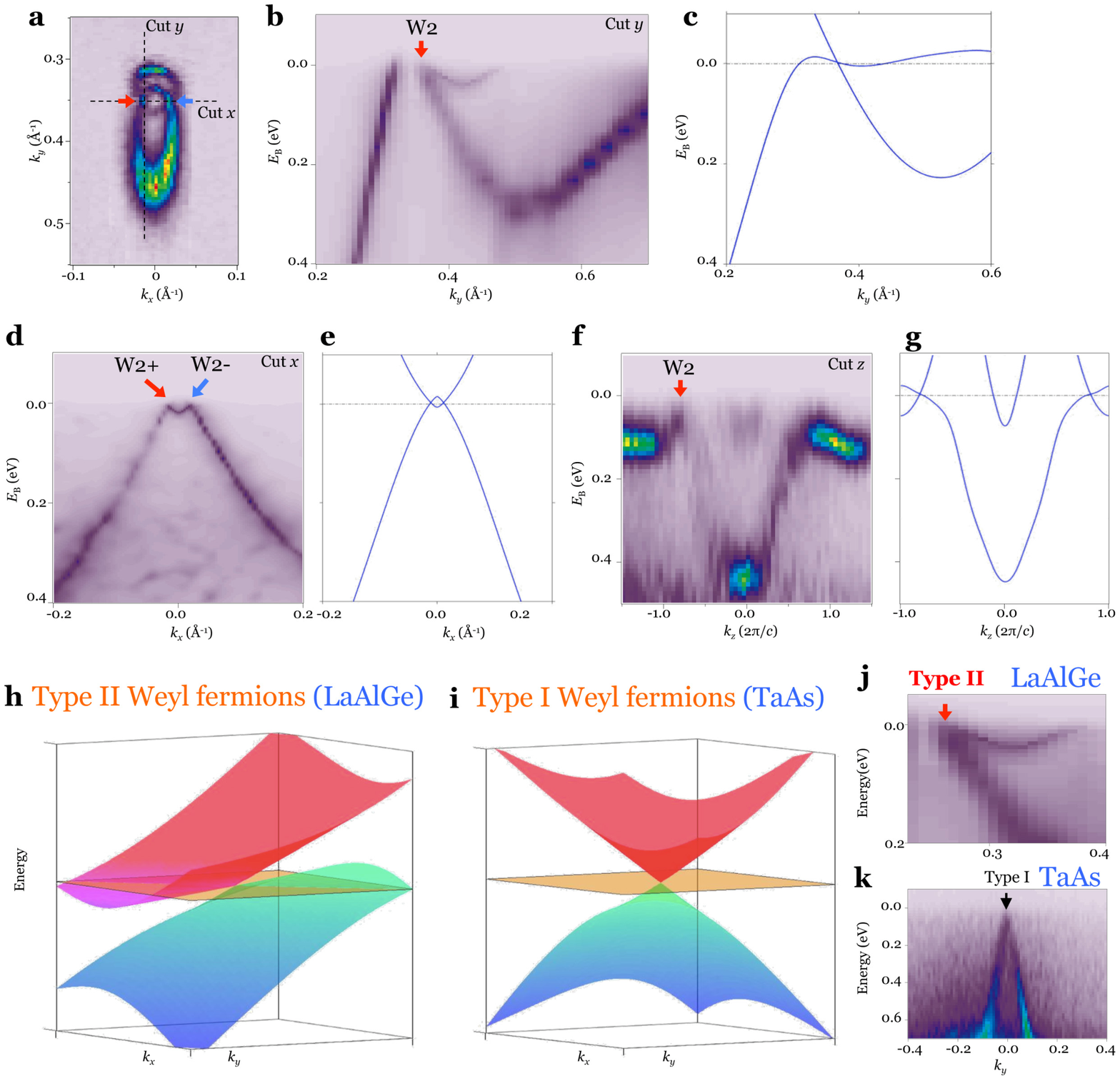}
\caption{
{\bf Type-II Weyl fermions in LaAlGe.} {\bf a,} SX-ARPES-measured $k_x-k_y$ Fermi surface map in the region marked by the green rectangle in panel {\bf c} of Fig. 2. {\bf b,} Measured and {\bf c,} calculated $E-k_y$ dispersion maps along the Cut $y$ direction denoted in {\bf a}, which clearly resolve the W2 type-II Weyl node emerged by the crossing of linearly dispersing hole-like and an electron-like Weyl cones. {\bf d,} Measured and {\bf e,} calculated $E-k_x$ dispersion maps for the direction along the Cut $x$ in {\bf a}. Here, the two type-II W2 Weyl cones with opposite chirality nodes are resolved simultaneously. The photon energy used in the measurements presented in {\bf a, b, d} is 542 eV. {\bf f,} Measured and {\bf g,} calculated $E -k_z$ dispersion map along Cut $z$, showing that the W2 Weyl cone also disperses}
\label{Bulk_disp}
\end{figure*}
\addtocounter{figure}{-1}
\begin{figure*}[t!]
\caption{linearly along the out-of-plane $k_z$ direction. {\bf h, i,} Schematics comparing type-II Weyl fermions observed here in LaAlGe, and type-I Weyl fermions observed previously in TaAs, respectively. {\bf j, k,} SX-ARPES-measured $E-k_y$ dispersion maps showing the difference between the type-II Weyl fermions of LaAlGe and type-I Weyl fermions of TaAs, respectively.}
\end{figure*}

\begin{figure*}[t]
\includegraphics[width=17cm]{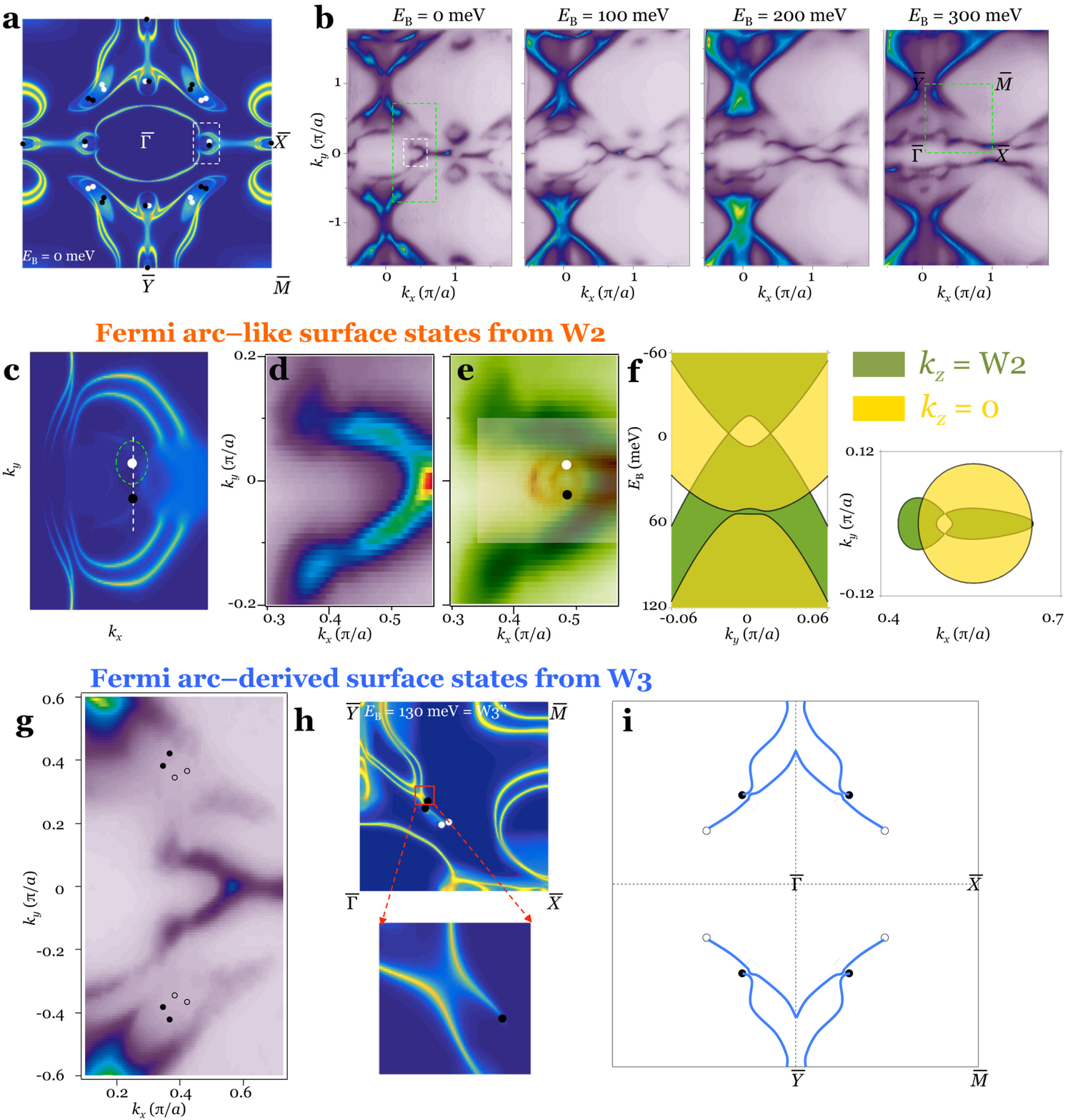}
\caption{ {\bf Fermi arc-like and Fermi arc-derived surface states in LaAlGe.} {\bf a,} First-principles calculated $k_x - k_y$ Fermi surface map of the (001) surface. {\bf b,} ARPES-measured $k_x - k_y$ Fermi surface map and three different constant binding energy contours at 100 meV, 200 meV, and 300 meV below the Fermi level. {\bf c,} First-principles calculated and {\bf d,} high-resolution ARPES-measured zoomed-in Fermi surface map along the $\bar{\Gamma}-\bar{X}$ direction around two W2 Weyl nodes, in the region indicated by the dashed white rectangle in the leftmost panel in {\bf b}. {\bf e,} Low-photon-energy ARPES Fermi surface map from {\bf d,} with the SX-ARPES Fermi surface map from Fig. 2{\bf d} overlaid on top}
\label{SS}
\end{figure*}
\addtocounter{figure}{-1}
\begin{figure*}[t!]
\caption{of it to scale, showing  the locations of the projected bulk Weyl nodes relative to the Fermi arc-like surface states that connect to the W2 Weyl nodes. {\bf f,} Schematics of an $E-k_y$ cut along the direction of the white dashed line in {\bf c} that shows how the Fermi arc connecting across the Weyl nodes gets overshadowed and masked by a trivial bulk band pocket located at $k_z = 0$. {\bf g,} Fermi surface map in a larger momentum region indicated by the dashed green rectangle in the leftmost panel in {\bf b}, which captures the momenta at which the W3'' Weyl nodes exist. {\bf h,} First-principles calculated $k_x-k_y$ constant energy contours, at the energy of 130 meV above the Fermi level corresponding to the energy of the W3'' node, in the region indicated by the dashed green rectangle in the rightmost panel in {\bf b}. The bottom panel is the zoomed-in region in the vicinity of a W3'' Weyl node, and shows the Fermi arc-derived surface states that connect to these nodes. The photon energy used in the measurements presented in {\bf b, d, g} is 50 eV. {\bf i,} Schematics illustration of the Fermi arc connectivity for the W3'' Weyl nodes.}
\end{figure*}

\end{document}